\documentclass[%
 reprint,
 floatfix,
 amsmath,amssymb,
 aps,
 prb,
]{revtex4-2}
\usepackage{enumitem}
\usepackage{varwidth}
\usepackage{tasks}

\usepackage{graphicx}
\usepackage{dcolumn}
\usepackage{bm}

\usepackage[T1]{fontenc}
\usepackage[margin=1in]{geometry}
\usepackage[utf8]{inputenc}
\usepackage[tone]{tipa}
\usepackage{tikz}
\usetikzlibrary{matrix,shapes,arrows,positioning,chains,calc}
\usepackage{adjustbox}
\usepackage{lipsum}
\usepackage{hyperref}
\hypersetup{colorlinks=true, citecolor=blue, urlcolor=blue, linkcolor=blue}
\newcommand{\asubsection}[2]{
\setcounter{subsection}{#1}
\addtocounter{subsection}{-1}
\subsection{#2}
}
\begin{document}

\preprint{APS/123-QED}

\title{Efficient determination of the true magnetic structure in a high-throughput ab initio screening: the MDMC method}

%

\author{Olga~Yu.~Vekilova}
\thanks{Corresponding author, olga.vekilova@mmk.su.se }
\affiliation{
Department of Materials and Environmental Chemistry, Stockholm University, 10691 Stockholm, Sweden
}


\date{\today}

\begin{abstract}
Finding the true magnetic structure at given external conditions is crucial for describing magnetic materials and predicting their properties. This is especially important for high-throughput screening of potentially good magnets that without adequate description of the magnetic structure may result in enormous waste of computational resources. I introduce a method, which accurately and efficiently treats magnetic ordering in the course of standard first-principles calculations. The method is suitable for all temperatures and is based on Monte Carlo (MC) technique. At high temperatures MC is coupled to ab initio molecular dynamics, therefore treating atomic vibrations and magnetic ordering simultaneously. The method takes care of the convergence to the true ground-state magnetic structure at 0 K, coupling between spins and atomic vibrations at high temperatures, and spin-spin interactions in non-Heisenberg systems, i.e. it naturally goes beyond the usual pair-like effective exchange interactions at all temperatures. I show that the method nicely reproduces the magnetic structures at low and high temperatures even on relatively small supercells and provides a ground for truly ab initio calculations of magnetic systems in high-throughput studies. In its general formulation the MDMC method deals with non-collinear magnetism, but its collinear version may be more suitable for fast high-throughput calculations. The examples of the application of the MDMC method addressed in this paper are $\alpha$-Mn; body-centered cubic Fe, which is ferromagnetic at low and paramagnetic at high temperatures; and MnB$_2$W$_2$, a new magnetic compound with non-trivial magnetism. 
\end{abstract}

\maketitle{}


\section{\label{sec:level1}INTRODUCTION
}

Magnetic properties of materials are of high interest to both fundamental science \cite{MagneticPropertiesBook2017_theoryandexp,Perversi2019,Recent_ferromagnet_NatureComm2018,PRL_Ni,NotRecent_Magnetism_MDMCneeded2009} and a wide range of applications, from permanent magnets for energy conversion to data storage \cite{COEY2020119,Matizamhuka2018,CUI2018118}. Even when magnetic properties are not the target in a particular study, one cannot properly describe physical or mechanical properties of a magnetic material at given external conditions (like temperature and pressure) without knowing its magnetic state. For example, changes in magnetic order (a typical example is a ferromagnetic (FM) to paramagnetic (PM) transition under heating) may deteriorate performance of magnetic devices or affect mechanical stability of industrially relevant magnetic alloys. 

In particular, in a search for new magnetic materials screening of potentially good permanent magnets is of great importance, as they are used in many green energy applications, such as wind turbines and electric cars. The most important characteristics of a good permanent magnet which can be predicted theoretically are the magnetic moment and magnetic ordering, magnetic anisotropy, and Curie temperature (T$_C$). In recent theoretical and experimental searches these properties were extensively studied \cite{vekilova_fe3sn, vekilova_smfe12,NIEVES2019188,Martinez_2019}. Many recent high-throughput approaches are designed in a way that during the first step an initial screening of the compounds with a proper total magnetic moment and magnetic ordering is done and then a more accurate and time-consuming study of magnetic anisotropy and T$_C$ is performed \cite{alena}. That is why it is important to have a method suitable for high-throughput studies, which is able to determine  the true magnetic structure and estimate the magnetic transition temperature with minimal efforts. 

A number of techniques to treat magnetic order as a function of external conditions have been developed over the years.   
Though they have been reported to successfully solve particular problems, there are clear drawbacks that hinder their usage for efficient predictions in large-scale simulations, such as high-throughput searches of new materials with given properties. Actually, even at 0 K temperature such searches rarely attempt to find the proper ground-state magnetic structure. They rather rely on the capabilities of standard density functional theory (DFT) to do so.  At best a limited number of magnetic ordering types, like the ferromagnetic and the simplest antiferromagnetic (AFM) ones, are tested. As I show below, such an approach is generally not valid when multiple energy minima are present in the space of magnetic solutions obtained in the DFT calculations. When temperature is raised, a typically used method either ignores coupling between the spin and vibrational degrees of freedom or it is so demanding  and time-consuming that its practical application is hindered by the currently available computational resources. Considering the existing schemes, I notice that often the magnetic structure at a particular temperature is calculated via a Monte Carlo simulation based on a Heisenberg-like Hamiltonian whose so-called exchange interaction parameters are estimated at 0 K. Even if the attempts to include atomic vibrations are made, as for instance, in the spin-space averaging approach \cite{Grabowski1,Grabowski2}, the atomic motion does not directly affect the magnetic structure.

An alternative is the spin-lattice dynamics (SLD), recent implementations of which can be found, e.g. in Refs.  \cite{Dudarev1,SPILADY,Dudarev3} and \cite{Perera1,Perera2}. An efficient usage of SLD cannot fully rely on ab initio calculations and requires adequate interatomic potentials for the magnetic and atomic degrees of freedom. Constructing those is particularly challenging for magnetic materials at high temperatures. 

A recently suggested atomistic spin dynamics with ab initio molecular dynamics (ASD-AIMD) method \cite{ASD_AIMD} tries to solve the problem by using model effective exchange interactions calculated from first principles for every atomic distribution obtained in the course of ab initio molecular dynamics (AIMD) simulations. Such an approach is a step forward towards better description of the spin-lattice dynamics. 
In particular, it has successfully described paramagnetic CrN \cite{ASD_AIMD}. However, ASD-AIMD relies on a model Heisenberg-like Hamiltonian with pair-wise effective exchange interactions.

\begin{figure*}[t]
\centering

\includegraphics[width=0.95\textwidth]{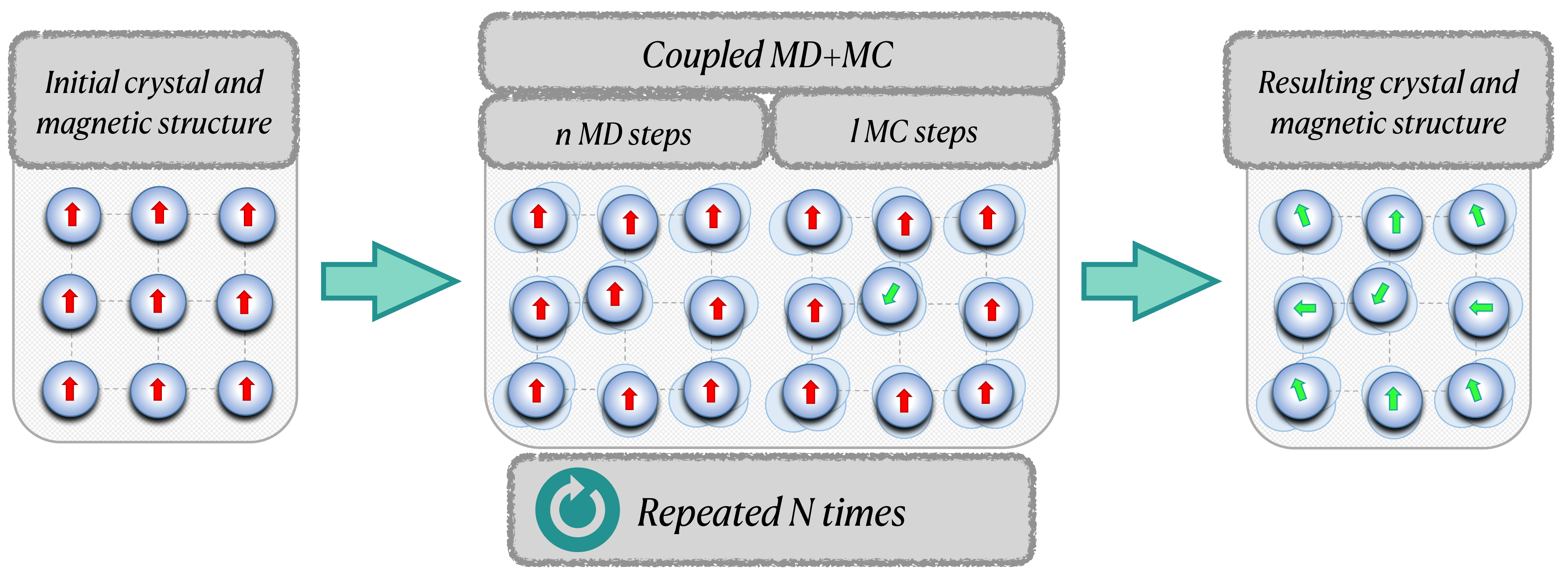}
\caption{Sketch of the MDMC procedure. $n$ is the number of MD steps and $l$ is the number of MC steps. Typically, 1 MD + 1 MC step are used. In the case of zero K temperature MD part is frozen but can be substituted by ionic/shape/volume relaxation DFT steps ($n>0$).}
\label{fig:method}
\end{figure*}

As was shown already in Ref. \cite{Ruban}, many materials (including the celebrated "Heisenberg system", body-centered cubic (bcc) Fe) turn out to be "non-Heisenberg", i.e. their effective exchange interactions strongly depend on temperature and therefore are very different in the high-temperature disordered (paramagnetic) and the low-temperature ordered (ferromagnetic, antiferromagnetic, or ferrimagnetic) states. This should in general be accounted for by recalculating effective interactions within ASD-AIMD. However, as was also shown in Ref. \cite{Ruban}, the multi-cite interactions can play as important role as the pair-wise ones. If this is the case, the Heisenberg Hamiltonian with just pair interactions might not be sufficient, and usually that is indicated by very large supercells used to reach the range of pair interactions that provide the desired convergence of the results.

The latter problem should in principle be tackled in the method suggested by Steneteg et al. in Ref. \cite{DLM-MD}. They introduced a so-called disordered local moment - molecular dynamics (DLM-MD) method, where MD runs in a maintained DLM state, so all the multisite interactions should be implicitly included up to the atomic distances within the size of the supercell. The reported results are in a reasonable agreement with experiment for the equation of state of CrN \cite{DLM-MD} and its elastic constants \cite{DLM_MD_elastic}. Nevertheless the method has other drawbacks. 

First, in the DLM-AIMD approach magnetic short and long range ordering effects are ignored. 
Formally that is in line with the purpose of the method to address an ideal paramagnetic state with complete disorder of magnetic moments. The DLM state is maintained at every moment in time, i.e. the total magnetic moment is forced to be zero. However, as one might expect, fluctuations of the total magnetic moment over the time keeping the average zero magnetic moment should be expected at any high but finite temperature. This is not catched by the DLM-MD, which in reality deals with systems at finite temperatures. 
Therefore the method's performance should deteriorate when the temperature is lowered. 

Further, the double spin flips that are introduced by DLM-MD at random atomic positions (a pair of random opposite spins are flipped at a time to maintain the DLM state), inevitably lead to situations where high-energy solutions with very low Boltzmann's factor are accepted, though they can hardly be realized at the addressed finite temperature. That is the phase space spanned during the MD run might not be the proper one, and a thermodynamic averaging of any property will likely be incorrect.  

As I show below the automatic search for the proper magnetic structure at 0 K and higher temperatures avoiding the mentioned above drawbacks can be naturally handled within the introduced here MDMC method. It is based on a smart Monte Carlo type simulation and at higher temperatures combines it with ab initio molecular dynamics simulations. In particular it finds the ground-state magnetic structure in a very efficient way. At high temperatures it is not limited to pair effective exchange interactions and allows for fluctuations of spins (their sizes and the magnetic order), spanning the phase space in a thermodynamically correct way, i.e. not accepting solutions with prohibitively low Boltzmann's factor. 

Generally, MDMC provides a framework for a thorough study of magnetic materials beyond the approximations used in standard or recently developed methods. However, in this paper 
I just aim at describing MDMC as the method that can relatively quickly and reliably find the magnetic state of a system at given external conditions, and therefore can be suitable for high-throughput studies. 
I also show some examples of the method applicability to the representative magnetic systems, such as bcc Fe at low and high temperatures, systems with complex magnetism, such as $\alpha$-Mn, and new compounds with interesting magnetic properties, such as MnB$_2$W$_2$. These examples illustrate that the method can be used for different purposes, in different cases of a priori unknown magnetic ordering, and for prediction of the magnetic structure  of new materials at any pressure-temperature conditions. 

\section{\label{sec:method}Method}


In the introduced here molecular dynamics - Monte Carlo (MDMC) method we run standard AIMD in the NVT or NPT ensemble (i.e. maintaining the number of atoms N, temperature T, and volume V or pressure P of the system) for a material, whose crystal structure can be described by a supercell with some initial atomic positions.  
At 0 K, in case the AIMD is used in its Newtonian formulation, the method effectively freezes up the MD part, substituting it by static DFT calculations, and quickly searches for the ground-state magnetic structure. The initial distribution of magnetic moments can be ferromagnetic, antiferromagnetic, DLM or anything allowed by the size of the chosen supercell and the number of magnetic atoms. 

As usual, AIMD (or static DFT at 0 K) runs within the spin-polarized density functional theory framework. That is a system of Kohn-Sham equations  is  solved self-consistently,  starting from  some input  distribution  of  atomic magnetic moments. This results in an output distribution of magnetic moments, which is not necessarily identical to the input one. Both sizes and directions of the magnetic moments can change. I notice that the solution of the Kohn-Sham equations for a spin-polarized system, obtained in standard first-principles codes, is not guaranteed to provide both ground state charge and magnetization density distribution ($\rho(\vec{r})$ and $m(\vec{r})$, respectively; $\vec{r}$ denotes the position in space)  for a particular set of atomic (nuclei) positions, $\vec{R_{i}}$, 
$i = 1,N$, where N is the total number of atoms in the cell). 

It can result in a metastable solution, which happened to be close to the starting point in space of $\rho$ and $m$ distributions  and provides a local minimum for the energy, which is the unique functional of  $\rho$ and $m$. The ground state $\rho$ and $m$ for the given set of $\vec{R_{i}}$ 
could in principle be reached if at every time-step t$_k$ we would try all distributions of initial magnetic moments possible for the given cell. That is certainly unfeasible for sufficiently large cells.

However, we can introduce a Monte Carlo-like step on top of the standard AIMD or zero-K static calculations. After time step t$_k$ in AIMD or the DFT ionic/shape/volume relaxation step $k$ at 0 K we flip one initial spin moment chosen at random, while keeping the set of $\vec{R_{i}}$, $i = 1,N$ 
intact, and recalculate the energy of the system. If it is lower than the one before the spin-flip, we accept the new magnetic configuration and proceed with the AIMD run or static relaxation according to the forces/stresses calculated for the new magnetic configuration. If the energy of the resulting structure is higher, we calculate Boltzmann's factor  $exp(-\Delta E/k_B \cdot T)$, where $\Delta E$ is the difference in energies between the magnetic configurations after and before the spin-flip and $k_B$ is the Boltzmann's constant, and compare it to a random number in the interval [0,1]. If the the random number is higher than the Boltzmann's factor, the new magnetic configuration is accepted.

By doing this for sufficiently long time, we cover the phase space of possible magnetic solutions directly coupled to the atomic vibrations in the case of AIMD. After a period of equilibration any statistics, which is of interest, such as magnetic moments, stresses etc. can be collected and analyzed. Obviously, at T = 0 K such an MC procedure leads to the true ground-state magnetic structure.

Let me formulate the suggested procedure step by step (see also its sketch in Fig. \ref{fig:method}). The course of action in an MDMC calculation is to run a standard AIMD or static DFT simulation extended at every $n$'th step (typically $n=1$ in AIMD and $n=1-10$ at 0 K) by $l$ Monte Carlo steps (typically $l$=1).
\begin{enumerate}[wide, label=\arabic*)]

  \item The starting magnetic configuration is set up. It can be ferromagnetic, antiferromagnetic, ferrimagnetic or DLM, non-collinear or collinear.
  \item $n$ MD time-steps (T $>$ 0 K) or  up to $n$ static relaxation steps in a DFT self-consistent run (T = 0 K) are performed.  Typically $n=1$ at T $>$ 0 K or in case of a fixed crystal structure at T = 0 K, and equal to the number of relaxation steps in case ionic positions and/or shape and/or volume of the cell are relaxed at T = 0 K.  The energy of the obtained structure at the $n$'th step, E$_1$, is recorded.
  \item In the structure obtained at step (2) an atom with non-zero on-site magnetic moment is chosen at random. The direction of the corresponding magnetic moment is changed by random angles in spherical coordinates, polar $\Delta \theta$ $\in$ $[0,\pi]$  and azimuthal $\Delta \phi$ $\in$ $[0, 2\pi]$. The window of allowed angles can be controlled and in the collinear version of the method $\Delta \theta$ $\equiv$ $\pi$ and accordingly $\Delta \phi$ $\equiv$ 0, i.e. a simple spin-flip is executed.
  \item The energy of the magnetic structure obtained in (3), E$_2$ is calculated. A comparison of E$_1$ and E$_2$ is made. If E$_2$ $\leq$ E$_1$, structure (3) is accepted. If E$_2$ $>$ E$_1$, then a random number $P$ $\in$ $[0, 1]$ is calculated and the decision to accept structure (3) or discard it and restore structure (2) is made based on
  
  \begin{equation}
  \begin{cases}
    (3)\ is\ accepted, & \text{if $P > e^{-\frac{E_{2}-E_{1}}{k_{B}T}}$}\\
    (2)\ is\ restored, & \text{otherwise},
  \end{cases}
\end{equation}

where $k_{B}$ is Boltzmann constant and $T$ is temperature.  

\item Step (4) is repeated $l$ times. Usually $l$=1.

\item Return to step (2).

\end{enumerate}

The whole procedure (2)-(6) is done N times, which is either a number of time-steps (at T $>$ 0 K) or DFT runs (at T = 0 K) sufficient to reach equilibrium magnetic configuration (at T $>$ 0 K) or the ground-state magnetic structure (at T = 0 K). Notice that  at T $>$ 0 K the system spans the phase space according to the thermodynamically correct probabilities at the given external conditions, so all the thermodynamic averages (like time-averaged on-site magnetic moments, stresses etc.) are readily available.

In the next sections I describe the computational details and show how the MDMC method works in practice.
I consider some of the revealing examples of the applications of the MDMC method. Out of many applications so far I choose those related to the high-throughput screening of magnetic materials. That is the primary interest lies in defining the magnetic ground-state structure and estimating magnetic order-disorder temperatures \cite{alena,vekilova_fe3sn,vekilova_smfe12,Martinez_2019,NIEVES2019188}.

\section{\label{sec:computational} Computational details}

 In the examples addressed in this paper I stick to the collinear version of the MDMC method. Though the MDMC method can be paired with any ab initio code, in the present work DFT calculations were done within the Projector-Augmented Wave method \cite{blochl1994} as implemented in Vienna Ab Initio Simulation Package \cite{kresse1996,kresse1996_2,kresse1999}. The recommended PAW potentials were used. 
 All the calculations were done with Perdew-Burke-Erzernhof exchange and correlation potential and energy \cite{Perdew1996}. Body-centered cubic iron was treated on a 54 atom $3\times	3\times	3$ cubic supercell. The energy cut-off for the plane-wave basis was set to 350~eV for 0~K case and to standard 268~eV for finite temperature simulations. Calculated equilibrium lattice parameter of 2.83~Å was used for the 0~K case. For the case of finite temperatures the lattice parameter was scaled depending on temperature as described in Sec. \ref {subsec:Fe_finite}. The initial magnetic moment on the Fe atom was $\pm3~\mu_{B}$ and 2.5~$\mu_{B}$ for zero T and finite T, respectively. The $2\times	2\times	2$ k-point Monkhorst-Pack (MP) grid \cite{Monkhorst-Pack} was used for all integrations over the Brillouin zone. The total number of MDMC steps was about 100 for the case of zero K and I tested up to $10 000-13 000$ MDMC steps for different temperatures in the case of finite temperatures.
 The AIMD simulations were performed in the NVT ensemble  with a time-step equal to 1 fs using the Nos\'e-Hoover thermostat.  $\alpha$-Mn was calculated for the 58 atom unit cell, with the energy cut-off 400~eV and integration over the Brillouin zone was done using the $\Gamma$-point. The initial distribution of magnetic moments was ferromagnetic, with $3~\mu_{B}$ on all the atoms. MnB$_2$W$_2$ was calculated for the $2\times	2\times	2$ 40 atom supercell, with the energy cut-off 600~eV and the $2\times	2\times	2$ k-point MP grid. The absolute value of  the initial  magnetic moment on the Mn atoms was set to $\pm5~\mu_{B}$. The number of MDMC steps was 200 for the runs starting from the DLM configuration and 1000 for the runs starting from the FM configuration. Calculations were done with full relaxation of volume, shape and ionic positions.

\section{\label{sec:0K}Finding the magnetic structure at 0~K}

Dealing with the systems at zero K no molecular dynamics is involved unless zero-point atomic motion is taken into account, which I do not address in this paper. The MDMC algorithm allows one to find the proper ground-state magnetic structure in an efficient way. In this case the introduced here method acts just like an extended MC simulation, where each MC step contains in itself a standard DFT run. 
I notice that the zero K case is of special importance to standard high-throughput screenings of new hard magnets \cite{alena}. 
It is currently a custom in such studies to choose the magnetic state of considered systems according to the literature or, in case such information is not available, to presume the ferromagnetic state \cite{alena}. Contrary to that, the MDMC approach quickly finds the ground-state magnetic structure. That in particular immediately discards the non-ferromagnetic materials, which are not suitable as hard magnets. 

\begin{figure}
\centering

\includegraphics[width=0.48\textwidth]{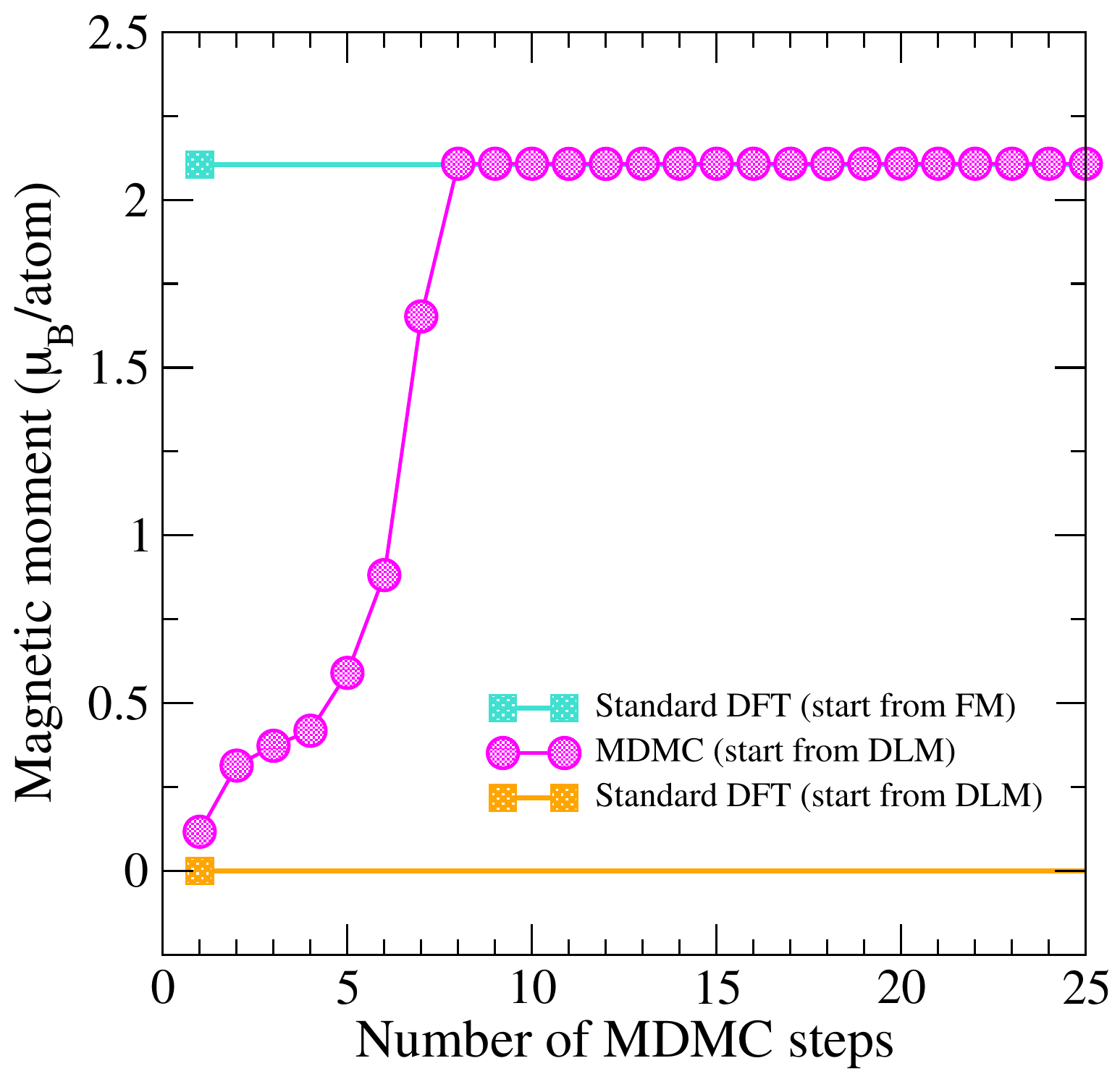}
\caption{Averaged magnetic moment per atom in bcc Fe at T = 0~K.
}

\label{fig:Fe0K}
\end{figure}

\asubsection{1}{\label{subsec:Fe0K}\emph{bcc Fe}}
Fe is a reference example of a magnetic material. 
At temperatures below 1185~K ($\alpha$-iron) and above 1663~K up to melting at 1811~K ($\delta$-iron) Fe has the body-centered cubic (bcc) structure. It is known to be ferromagnetic below 1043~K with high magnetic moment per Fe atom ($\sim$ 2.12 $\mu_{B}$) \cite{Han}. I studied the behavior of bcc $\alpha$- iron, space group $Im\bar{3}m$, at zero K. 
The purpose was to illustrate the process of magnetic ordering in a supercell geometry and difficulties with its determination using standard DFT approaches. 
As a starting magnetic state I used a disordered local moment state with spin moments on all atoms  equal to $\pm$3~$\mu_{B}$ with the total moment equal to zero. It is well known that the collinear DLM state usually provides a good description of the non-collinear PM state. The result was compared to the direct calculation of the proper ferromagnetic (FM) structure.
\begin{figure}
\centering

\includegraphics[width=0.48\textwidth]{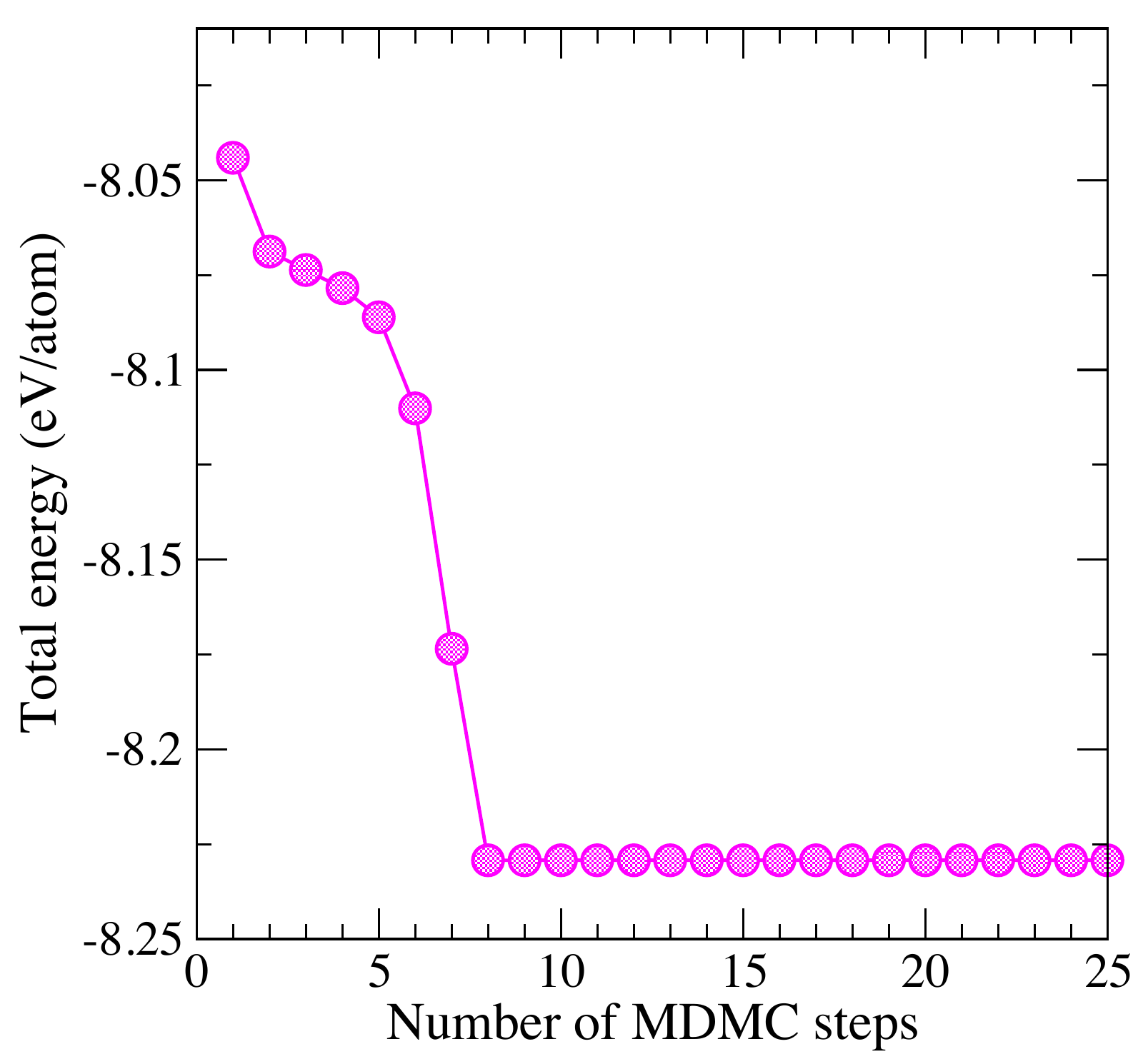}
\caption{Convergence of the total energy per atom of bcc Fe in a 54 atom supercell at T = 0~K in the MDMC method. 
}

\label{fig:Fe_energy}
\end{figure}

Obviously, in a standard DFT calculation for the bcc unit cell with one Fe atom the ferromagnetic state would be easily achieved as the only possible solution. This ease is often erroneously assumed to propagate to the calculations of larger supercells.
That is a wrong assumption. A standard DFT calculation 
with typical convergence criteria for the total energy 
may converge and finish without changing the starting magnetic structure. 
That is a standard DFT calculation may never end up in the proper magnetic state if no information about the distribution of magnetic moments in the system is presumed. Though this fact should not surprize an experienced DFT user, it is overlooked rather often. When the problem is understood, usually two possible cases are calculated, typically the FM and the simplest antiferromagnetic (AFM) orderings, and their energies are compared. The system with a lower energy is picked up as the equilibrium magnetic state. This approach obviously works in simple cases, especially when the system is ferromagnetic. However, if a more complex, say antiferromagnetic, ferrimagnetic or non-collinear ordering is the true magnetic structure at given external conditions,  it may never been revealed in such calculations. 

The common wisdom tells us that bcc Fe should be a simple case with the well-defined FM ordering, corresponding to a sufficiently deep minimum on the energy surface in the space of possible $\rho(\vec{r})$ and $m(\vec{r}$). Indeed the FM initial distribution of the magnetic moments is conserved in the course of the self-consistent DFT run for the considered supercell (turquoise square in Fig. \ref{fig:Fe0K})). However, the calculation of the same supercell in the DLM initial distribution of the magnetic moments does not automatically fall into the FM one (orange square in Fig. \ref{fig:Fe0K})). It stays DLM, clearly indicating the presence of multiple minima on the energy surface.

When I apply the MDMC method  and start from the DLM state, it 
quickly evolves into the ferromagnetic state after just 8 MDMC steps (see the purple line in Fig. \ref{fig:Fe0K}). 
It is worth noticing that the MDMC method efficiently uses the effect of what I call induced spin-flips, i.e.  the ability to flip more than one atomic magnetic moment in the course of the self-consistent DFT calculation at an MDMC step, which itself flips just one spin at a time. Fig. \ref{fig:Fe_energy} shows how the energy converges during the MDMC run. As expected, the MDMC method exhibits a steep energy decrease towards the ground state. After the ground state energy is achieved and the proper magnetic distribution is reached the system will stay in a current state, no matter how many new MDMC steps will be done. Therefore the convergence of the method into the ground state is easily monitored.

\asubsection{2}{\label{subsec:Mn}\emph{\texorpdfstring{$\alpha$-Mn}{Lg}}}

In some cases, especially when we aim at a theoretical prediction of new structures with complex magnetism, it is not enough to just determine whether the system is ferromagnetic or not. Even knowing the proper total magnetic moment of the system, some important features, like the type of the AFM ordering and values of magnetic moments on different atoms might be missing, leaving the answers to experimental studies. The MDMC method straightforwardly overcomes these issues. For an illustration I consider the second example, $\alpha$-manganese in its ground state, space group $I\bar{4}3m$. Mn has unique  structural and magnetic properties. While all other $3d$ metals attain simple crystalline structures with $1-2$ atoms/cell and simple ferromagnetic, antiferromagnetic or non-magnetic configurations, the ground-state $\alpha$-Mn adopts a very complex structure with 58 atoms in the cubic cell with four types of Mn positions, each with different magnetic moments in a range from 0 to $\pm 3.3~\mu_{B}$ according to a number of theoretical and experimental works (see Ref. \cite{Bradley,Hobbs} and Refs. therein). The magnetic structure is a complex antiferromagnet and in fact slightly non-collinear, however a reasonable collinear approximation to it is usually sufficient to treat the energetics of the system \cite{Hobbs}. As I stick to the collinear version of  the MDMC method in  this paper, I exploit this fact and consider the system at the experimental equilibrium volume, allowing for ionic relaxations in the course of the MDMC run, with maximum $n=10$. The latter is sufficient to get the final structure with targeted forces on all the atoms less than $5\cdot10^{-3}$~eV/Å.

\begin{figure}!
\centering

\includegraphics[width=0.47\textwidth]{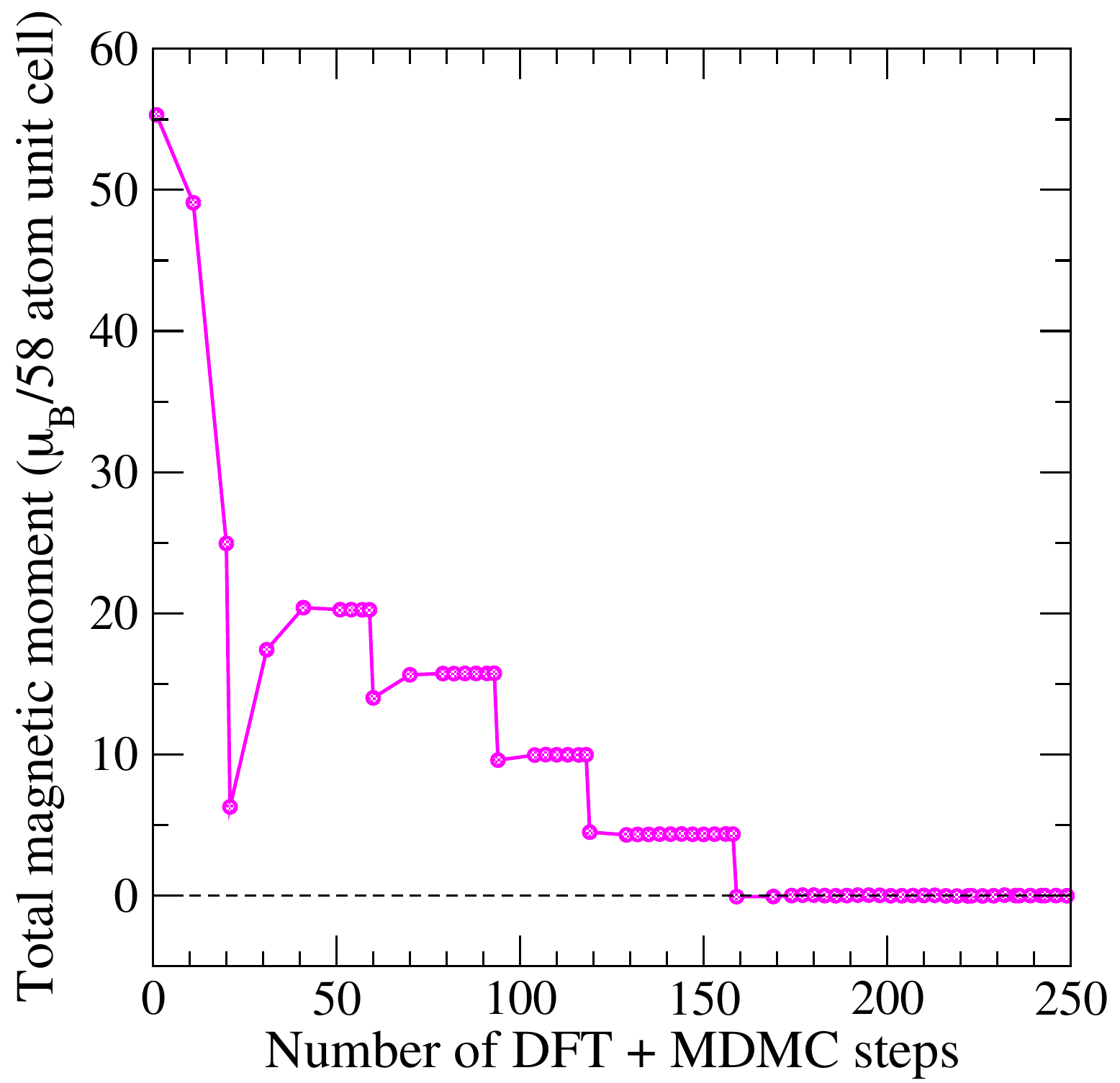}
\caption{Convergence of the total magnetic moment of $\alpha$-Mn in a 58  atom cell at T = 0~K in the course of the MDMC run.  Each circle represents 1 MDMC step. Lines between the circles correspond to the DFT ionic relaxation steps. Maximum number of 10 DFT ionic relaxation steps was allowed for each MDMC step.}

\label{fig:Mn0K}
\end{figure}

\begin{figure}
\centering

\includegraphics[width=0.48\textwidth]{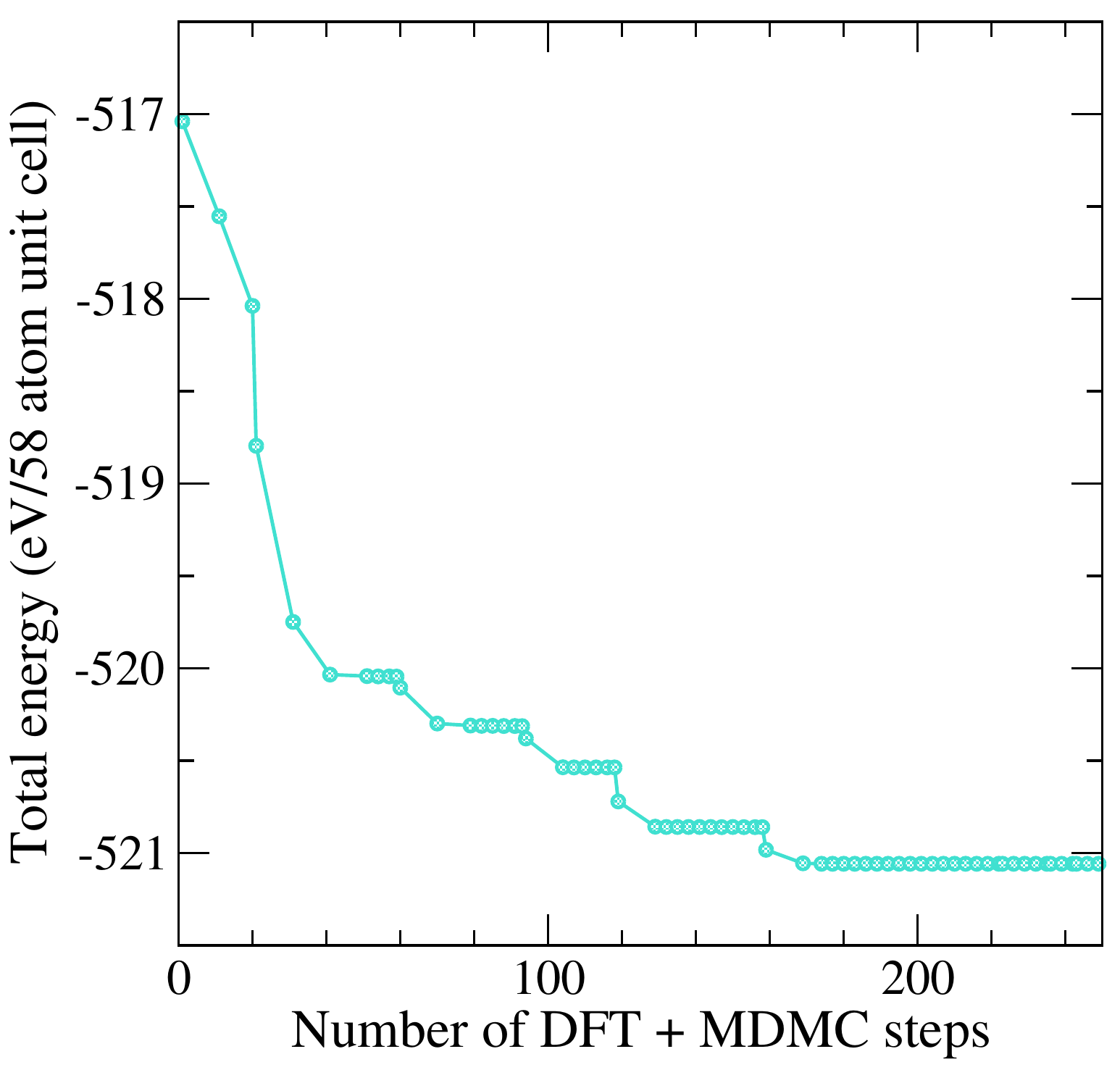}
\caption{Convergence of the total energy per atom of $\alpha$-Mn in a 58 atom cell at T = 0~K in the course of the MDMC run. Each circle represents 1 MDMC step. Lines between the circles correspond tothe DFT ionic relaxation steps. Maximum number of 10 DFT steps was allowed for each MDMC step. }

\label{fig:Mn_energy}
\end{figure}

I deliberately start from the "wrong" ferromagnetic structure  with the moments on all 58 atoms equal and set to 3~$\mu_{B}$. The collinear version of the MDMC code finds the proper collinear ground state 
within just 38 MDMC steps (159 DFT ionic relaxation steps in total), see Figs. \ref{fig:Mn0K} and \ref{fig:Mn_energy} for the convergence of the total magnetic moment and the energy, respectively. 

\begin{figure}
\centering

\includegraphics[width=0.45\textwidth]{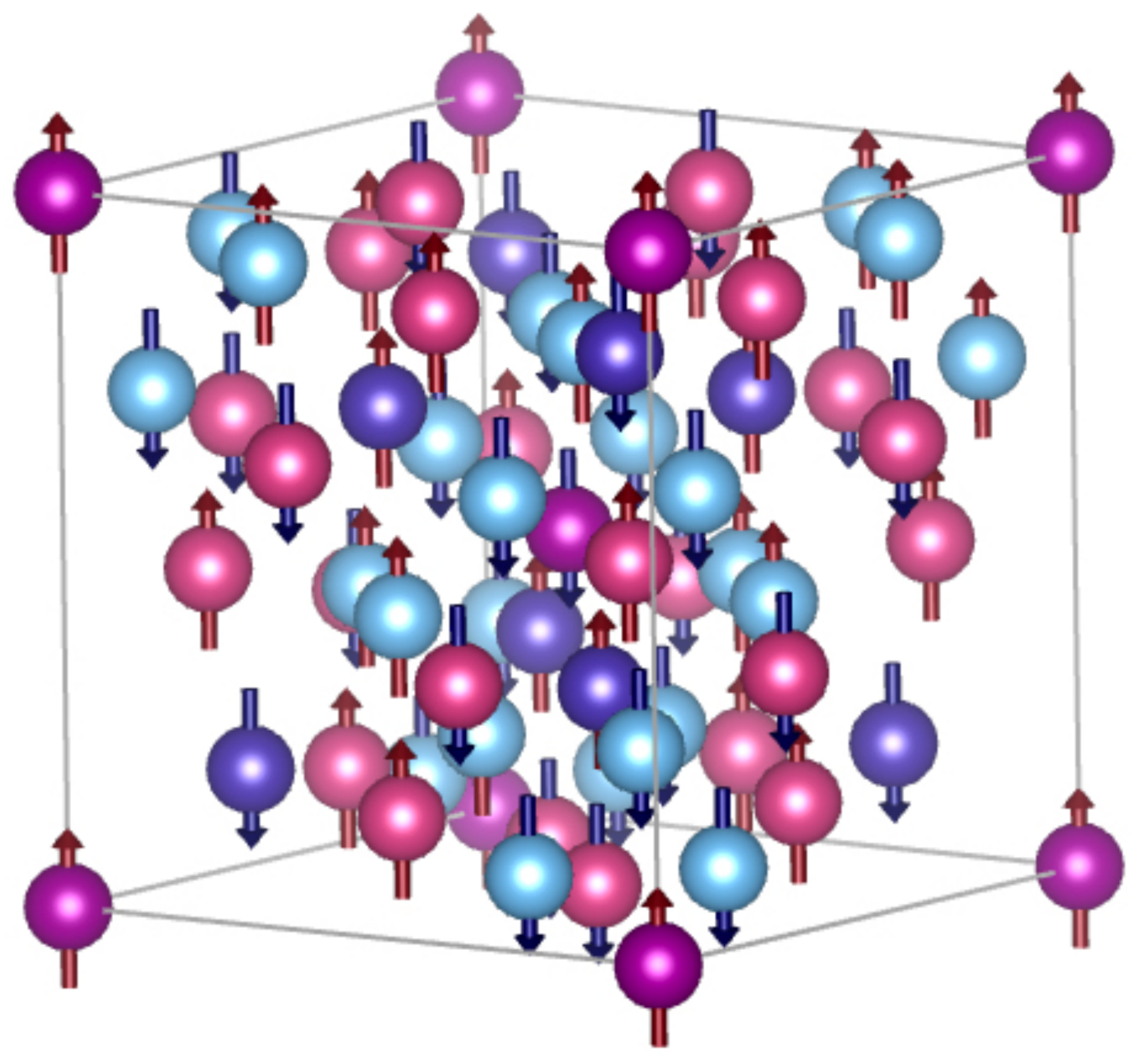}
\caption{Collinear version of the $\alpha$-Mn magnetic structure predicted by the MDMC method. Violet atoms represent the Mn sublattice with m = 3.4~$\mu_{B}$ (type I), dark blue correspond to m = 2.9~$\mu_{B}$ (type II), light blue have m $\sim$0~$\mu_{B}$ (type III), while purple are with m = 2.1~$\mu_{B}$ (type IV).}

\label{fig:Mn_str}
\end{figure}

The converged MDMC total magnetic moment in the 58 atom unit cell is zero, as expected in the antiferromagnetic $\alpha$-manganese. Strict division into 4 groups of Mn atoms with moments equal to $\pm 3.4~\mu_{B}$, $\pm 2.9~\mu_{B}$, $\pm 2.1~\mu_{B}$, and around $\pm 0~\mu_{B}$ is obtained. The final distribution of magnetic moments is illustrated in Fig. \ref{fig:Mn_str}.  These results are in good agreement with those reported in the literature \cite{Bradley,Hobbs}. This indicates that the MDMC method properly describes even systems with complex magnetic structures and allows us to study in an efficient manner rather difficult cases with no presumed information on the magnetic structure.

\asubsection{3}{\label{subsec:MnBW}\emph{\texorpdfstring{MnB$_2$W$_2$}{Lg}}}
The structure of the MnB$_2$W$_2$ compound was experimentally discovered in 1971 \cite{Telegus197152} and since then it was believed to be non-magnetic according to both, experiment and theoretical simulations \cite{W2MnB2_nonmagnetic_firstprinciples}. Recently the MnB$_2$W$_2$ structure with space group $P4/mbm$ was selected as a possible permanent magnet in a high-throughput and data-mining study  \cite{alena}. No magnetic structure, however, was determined, and the system was reported as non-magnetic \cite{alena}. 
Therefore I apply here the MDMC method to find out the true magnetic structure of this compound. This compound can also serve as an illustration of the predictive power of the MDMC method
for complex compounds containing both magnetic and non-magnetic elements. 

\begin{figure}
\centering

\includegraphics[width=0.48\textwidth]{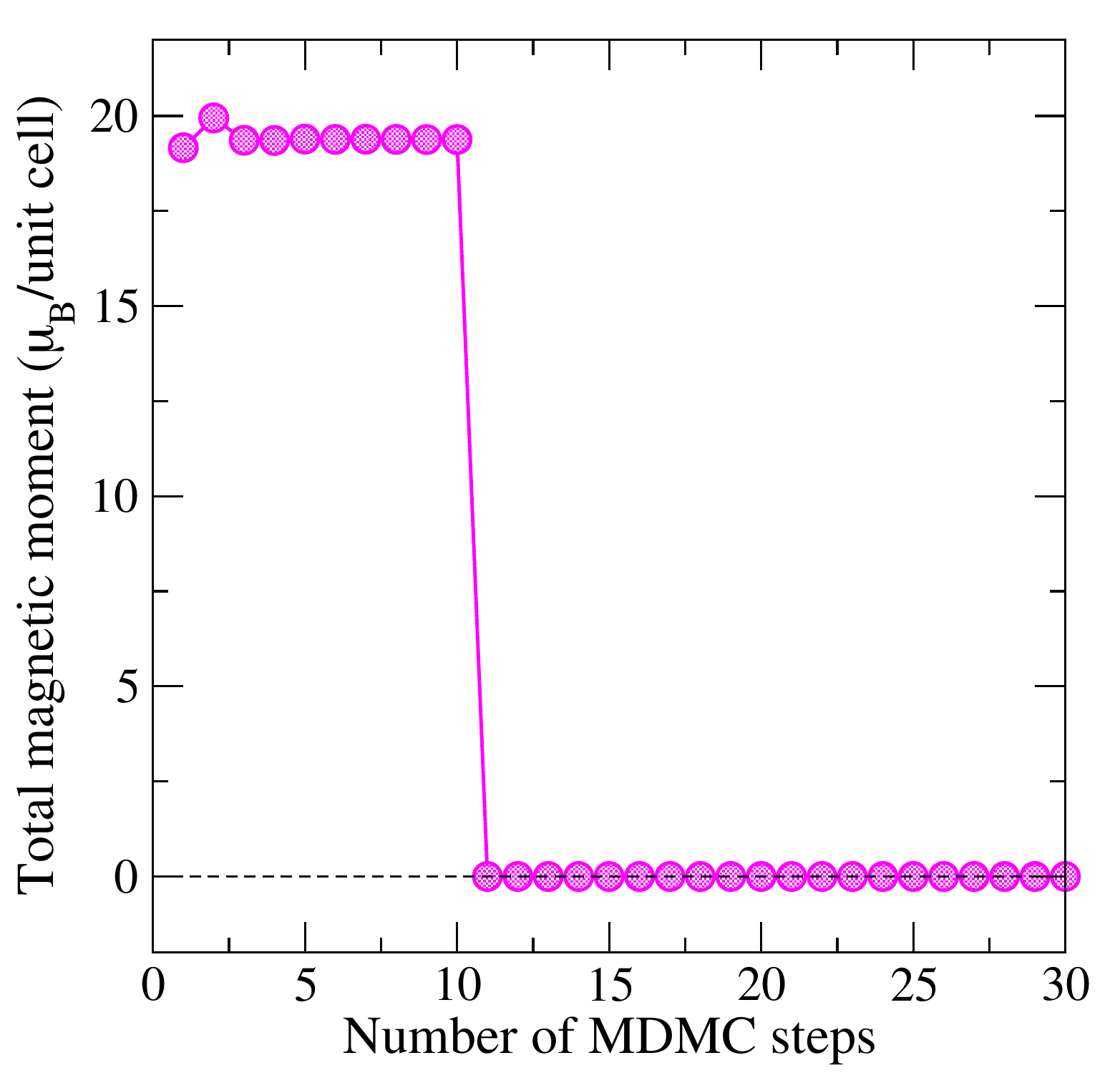}
    \caption{Convergence of the total magnetic moment from the initial FM into the final AFM state in MnB$_2$W$_2$ at T = 0~K. Only the first 30 MDMC steps are shown. However, the system remains converged in this magnetic state during 200 following steps. Notice that in the considered cell with 8 magnetic Mn atoms the initial total moment of the FM state was set to  $m = 5\times 8 = 40\mu_B$, however, after the first MDMC step it drops to  $\sim$19.4 $\mu_B$.}

\label{fig:MnWB1}
\end{figure}

\begin{figure}[t!]
\centering
\includegraphics[width=0.48\textwidth]{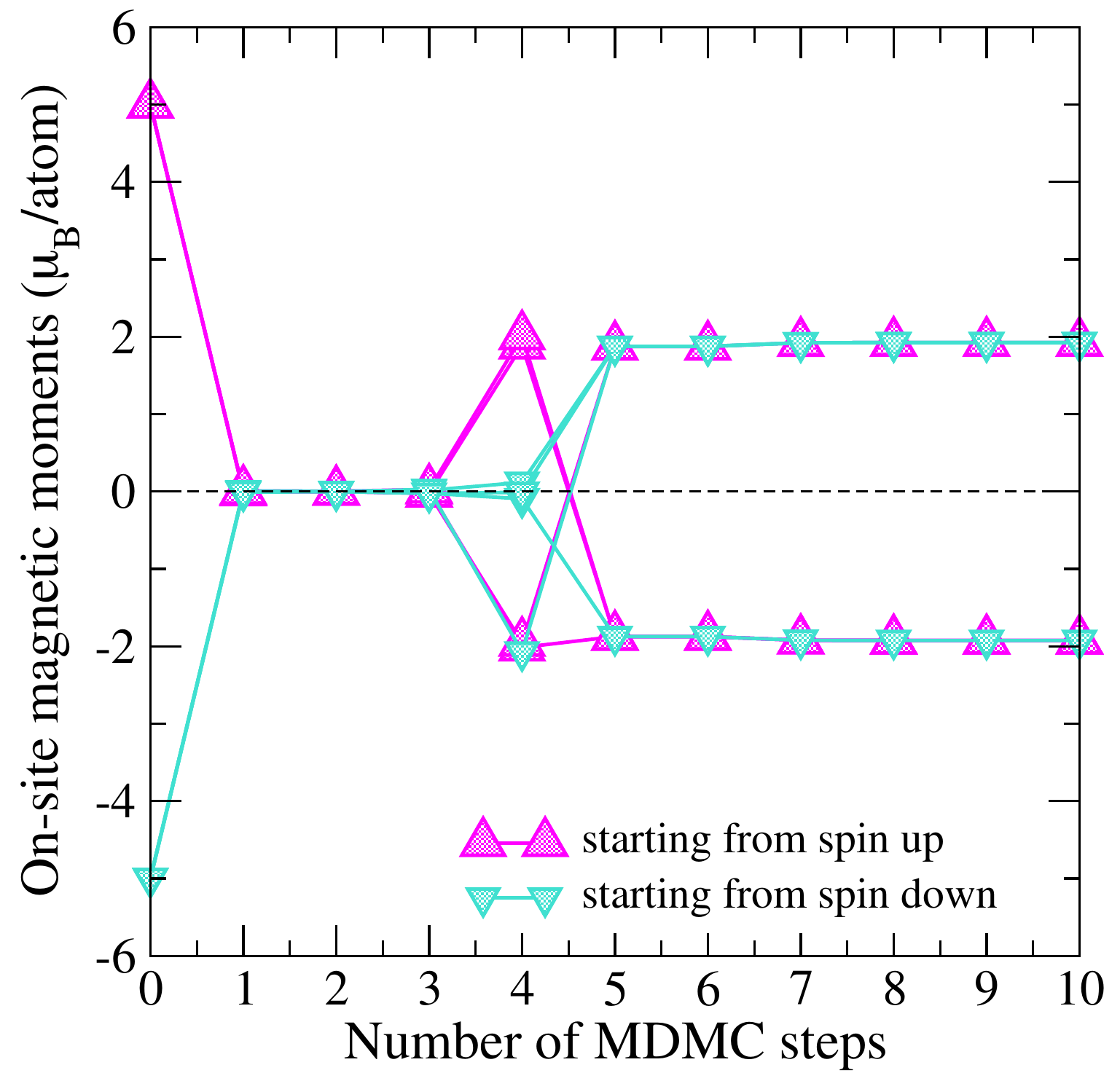}   
\caption{Convergence of magnetic moments on 8 Mn atoms from the initial DLM state into the AFM state in MnB$_2$W$_2$. MDMC step 0 corresponds to the initial magnetic structure with 4 spin up and 4 spin down magnetic moments on the Mn atoms with the size of $\pm5~\mu_{B}$. Magnetic moments on 8 Mn atoms naturally split into 2 groups. Only first 10 MDMC steps are shown. However, it remains converged during 200 following steps.}

\label{fig:MnWB2}
\end{figure}

For this example I chose to run calculations in the course of the collinear version of the MDMC method with the preset symmetry but no presumed volume or shape of the cell. That is MDMC automatically couples DFT relaxation of shape, volume, and ionic positions with the search for the ground-state magnetic structure. This is an option that appears especially useful in a high-throughput screening, in case the information on the crystal structure is insufficient. $n$, as defined above in Sec. \ref{sec:method} was set to 10, i.e. up to 10 steps of volume, shape, and ionic relaxations were allowed between the MDMC spin-flips. Two starting points were considered, the ferromagnetic and DLM structures. The initial value of the magnetic moment on each Mn atom was chosen as $m = \pm 5~\mu_{B}$. i.e. close to the values corresponding to the high-spin Mn configuration.   
Starting from the ferromagnetic state, the development of magnetic moments in the course of the MDMC run is rather simple (see Fig. \ref{fig:MnWB1}). The moments drop to the value $m \sim 1.92~\mu_{B}$, which corresponds to the low-spin Mn configuration, and then spin-flips lead to an antiferromagnetic ground-state structure, which is found within 11 MDMC steps (i.e. 11 spin-flips) (Fig. \ref{fig:MnWB1}). The ground-state structure 
turns out to have a layered distribution of moments along the c axis. This is the so called A-type antiferromagnetic ordering. 
When starting from a disordered local moment distribution  the system evolves into the same AFM solution, as obtained in the MDMC run departing from the FM state, within 8 MDMC steps (i.e. 8 spin-flips). 
The development of magnetic moments on each of the eight Mn atoms in the supercell of  MnB$_2$W$_2$ in this case 
is rather revealing. It is shown in Fig. \ref{fig:MnWB2}.
Right at the first MDMC step all the moments on Mn drop to nearly zeroes, and they are, of course, zeroes on all other atoms. However, from MDMC step 4 one can see that some of the moments get new value of $\sim\pm 2~\mu_{B}$, while other remain nearly zeroes. Finally after MDMC step 6 final AFM distribution with $m = \pm 1.92~\mu_{B}$ 
is attained, and magnetic moments are not changing after MDMC step 8. At this point the global total energy minimum is reached and for the following steps (200 MDMC steps were tested)  magnetization of the system does not change. 

It is worth noticing that the initial disordered distribution of magnetic moments with zero total moment in the case of the start from the DLM state evolves into the structure with the same total magnetic moment but completely different magnetic order. This in particular shows that the MDMC method successfully discerns different types of  AFM 
orderings.

\begin{figure*}[t!]
\centering

\includegraphics[width=0.95\textwidth]{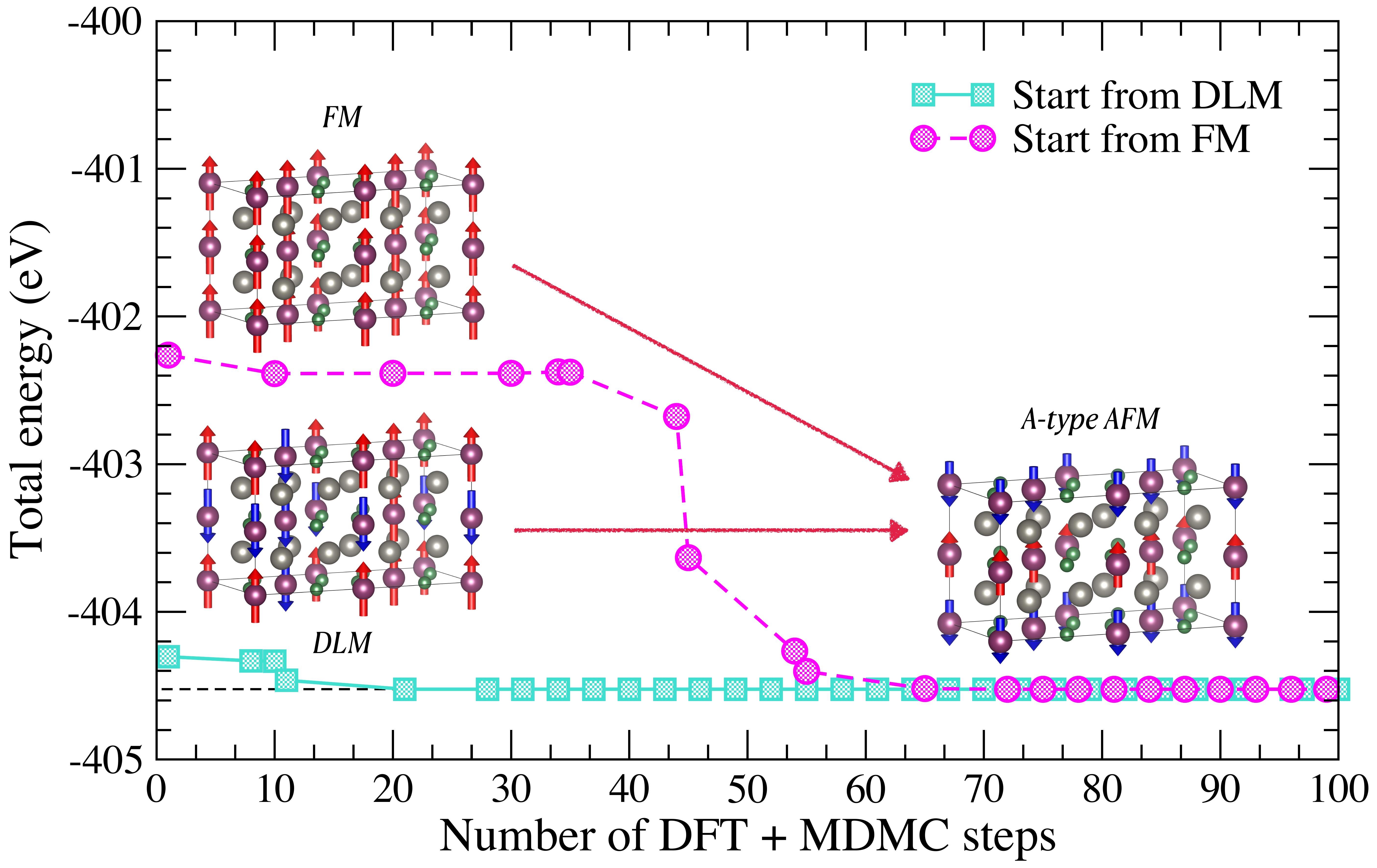}
\caption{Total energy convergence of MnB$_2$W$_2$ in the course of the MDMC runs. The transition of the DLM and FM  magnetic distributions into ordered A-type AFM in  MnB$_2$W$_2$ system at T = 0~K is shown together with the magnetic structure plots. Red and blue arrows in the magnetic structures represent spins up and down on the Mn atoms. W atoms are shown with grey spheres and B atoms are shown with small green spheres. Each circle on the energy curves represents 1 MDMC step. Lines between the circles correspond to the DFT ionic+shape+volume relaxation steps. Maximum number of 10 such DFT steps were allowed for each MDMC step.} 

\label{fig:MnWB3}
\end{figure*}

In Fig. \ref{fig:MnWB3} evolution of the total energy during the transition from both the DLM and FM into the A-type AFM state is shown together with the corresponding magnetic structures. 
An important message, which cannot be underestimated, is that  a standard DFT calculation of the DLM state would give a clear non-magnetic solution, in line with a 50 year old assumption in the literature. Indeed, as one can easily conclude from Figs. \ref{fig:MnWB3} and \ref{fig:MnWB2}, the energy of the calculated "non-magnetic" solution is lower than the energy of the ferromagnetic state. 
However, the energy surface in the $\rho$ and $m$ space turns out to be so complex that there is a non-trivial path from the non-magnetic solution into the well-defined antiferromagntic one, which is the true ground state. Obviously, such an unexpected behavior of this complex compound indicates that a method like MDMC is particularly needed to avoid wrong conclusions in high-throughput screenings of magnetic materials.


This example shows that even for a complex compound the results of the MDMC approach do not depend on the initial distribution of magnetic moments, and therefore it appears suitable for systems with a priori unknown magnetic properties and/or structural parameters. 

\section{\label{sec:finiteT}Finding the magnetic structure at high temperatures}

For a high-throughput screening of magnetic materials the point of highest importance is a reasonable estimation of the magnetization as a function of temperature and the order-disorder magnetic transition temperature. For example, a good hard magnet has to be not only ferromagnetic, but also its Curie temperature should be at least of order of 400 K. Therefore one should be able to find the magnetic structure at high temperatures to make reasonable predictions. 
Formally that could be done via calculations of the free energy instead of the total energy, as it is done at 0 K. 
In reality to calculate free energies of high-temperature magnetic systems without severe approximations is a formidable task. 

The MDMC method at high temperatures relies on the ab initio molecular dynamics and builds the magnetic relaxation cycle into a regular AIMD run, coupling the magnetic degree of freedom with the vibrational one in a fully ab initio way. The magnetic phase transformations are naturally observed in the course of MDMC calculations on a grid of temperatures, so the transition temperature can be estimated
without calculations of the free energy, solely on the basis of the magnetization vs temperature dependence. It is important that at a high temperature magnetic structure is generally no longer static but essentially dynamic. Both sizes and directions of magnetic moments on magnetic atoms may significantly fluctuate as a function of time depending on the atomic surroundings at a particular time-step. i.e. on the magnetic moments on the neighboring atoms and momentarily distances to them at a particular time-step.
Therefore converged time-averaging of the physical properties of
interest is usually required.

\asubsection{4}{\label{subsec:Fe_finite}\emph{bcc Fe at finite T}}

I chose as an example bcc Fe, which was already considered in Sec. \ref{sec:0K} for the case T = 0~K. 
As mentioned above, experimental Curie temperature of bcc iron is 1043~K. At this temperature the FM state transforms into the paramagnetic (PM) one, which can be reasonably approximated by the DLM state. In fact, high-temperature bcc Fe is a test-case for any method claiming to be able to predict the magnetic transition temperature. There are many publications on the FM-to-PM transition in bcc Fe, some recent examples can be found in Refs. \cite{T_C_Fe_bcc_YinPRB2012_largecell,Dudarev3,Han,T_C_Fe_bcc_calc_HellsvikPRB2019_largecell}. A number of them report transition temperatures in good agreement with experiment, others provide worse agreement. 
More important is that most of those calculations are either very involved \cite{Han} or need very large supercells to obtain  reasonable results \cite{T_C_Fe_bcc_YinPRB2012_largecell,T_C_Fe_bcc_calc_HellsvikPRB2019_largecell}. 
Let me show how the MDMC method performs in predicting magnetic transitions if a reasonably small supercell is considered. That is whether it may be of use in high-throughput screenings of magnetic materials when fast calculations are the requirement. In this section I am considering results for the $3\times	3\times	3$ 54 atom cubic supercell, as described in Sec. \ref{sec:computational}.

I again underline that the essential difference of the proposed method is that in MDMC the atomic vibrations with any degree of anharmonicity are directly coupled to the magnetic degree of freedom and there is no limitation to Heisenberg-like interactions, i.e. all the pair and multisite effective exchange interactions are implicitly included up to the range provided by the size of the supercell and are temperature-dependent. Accordingly, it is a natural expectation that smaller supercells could be used in the MDMC method. 

To see whether the magnetic transition in bcc Fe is reproduced I first chose a set of temperatures between 300 and 2000~K and initially used lattice parameters interpolated according to the experimental data between 298 and 1800~K  \cite{Lattice_expansion} to save time needed for the search of equilibrium volumes corresponding to time-averaged zero pressures. 

\begin{figure}
\centering

\includegraphics[width=0.48\textwidth]{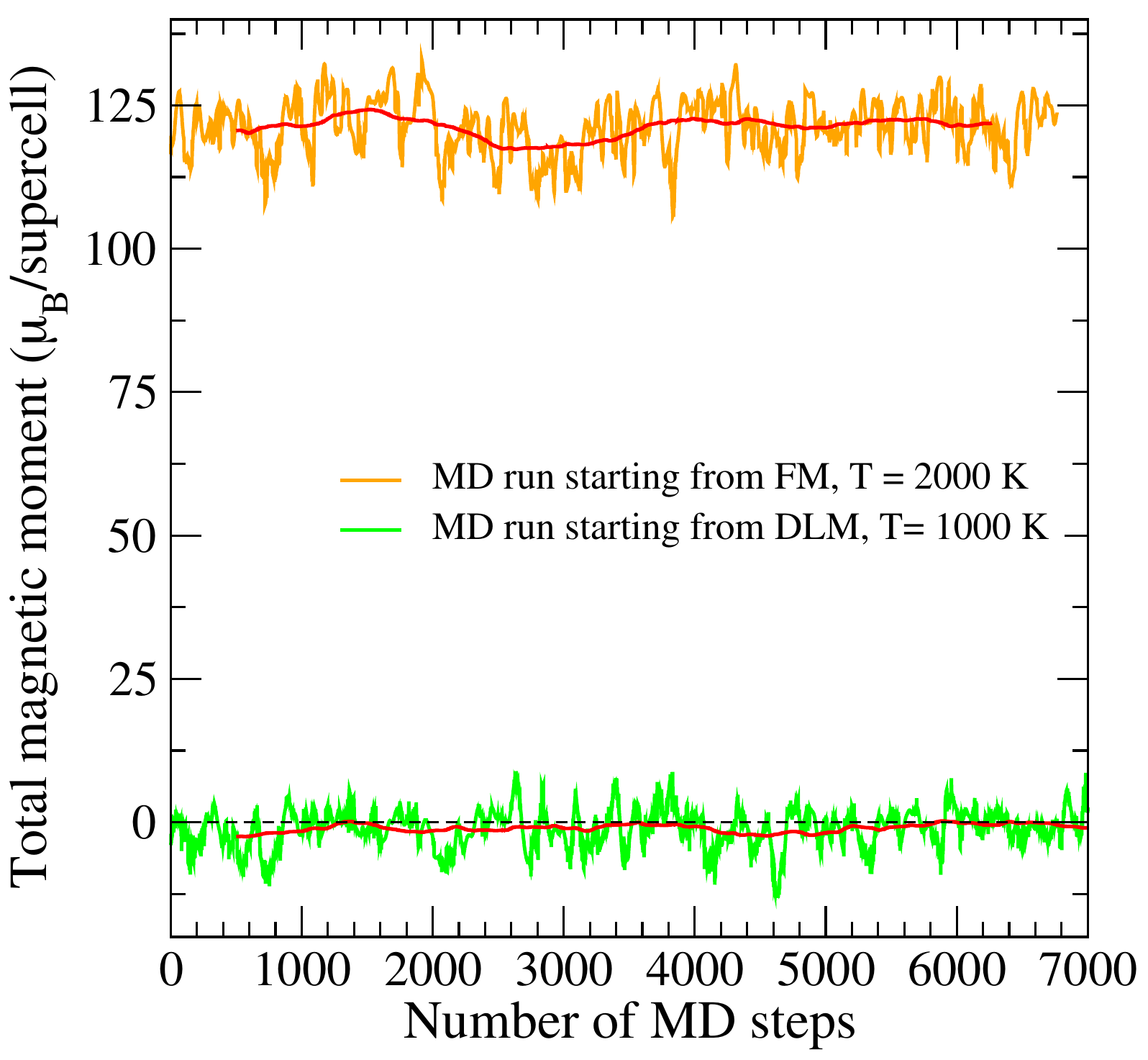}
\caption{Standard MD run without MDMC technique. Magnetic moment of  bcc Fe at T = 2000~K (above Curie temperature) and at T = 1000 (below Curie temperature). Running averages are shown with red lines. }

\label{fig:Fe_nomdmc}
\end{figure}

Two of these points, at T=1000 and 2000 K, were first checked without the MDMC method in a standard AIMD run  (Fig. \ref{fig:Fe_nomdmc}). 1000~K is expected to be just below and 2000~K should be well above the transition.  Accordingly, the case T = 1000~K should be close to a ferromagnetic solution, and the case T = 2000~K should exhibit clear paramagnetic behavior. Therefore I started the 1000~K run from a "wrong" DLM state, and the 2000~K run from the "wrong" ferromagnetic one, to double-check that a standard AIMD preserves the initial magnetic structures. Indeed, as one can see from Fig. \ref{fig:Fe_nomdmc}, AIMD starting from the FM state stays in it at 2000~K for the whole run (7 ps). The same happens to the system at 1000~K if one starts the AIMD simulation from the DLM state. It stays in a perfect DLM state, not at all approaching the FM ordering.

\begin{figure}
\centering

\includegraphics[width=0.48\textwidth]{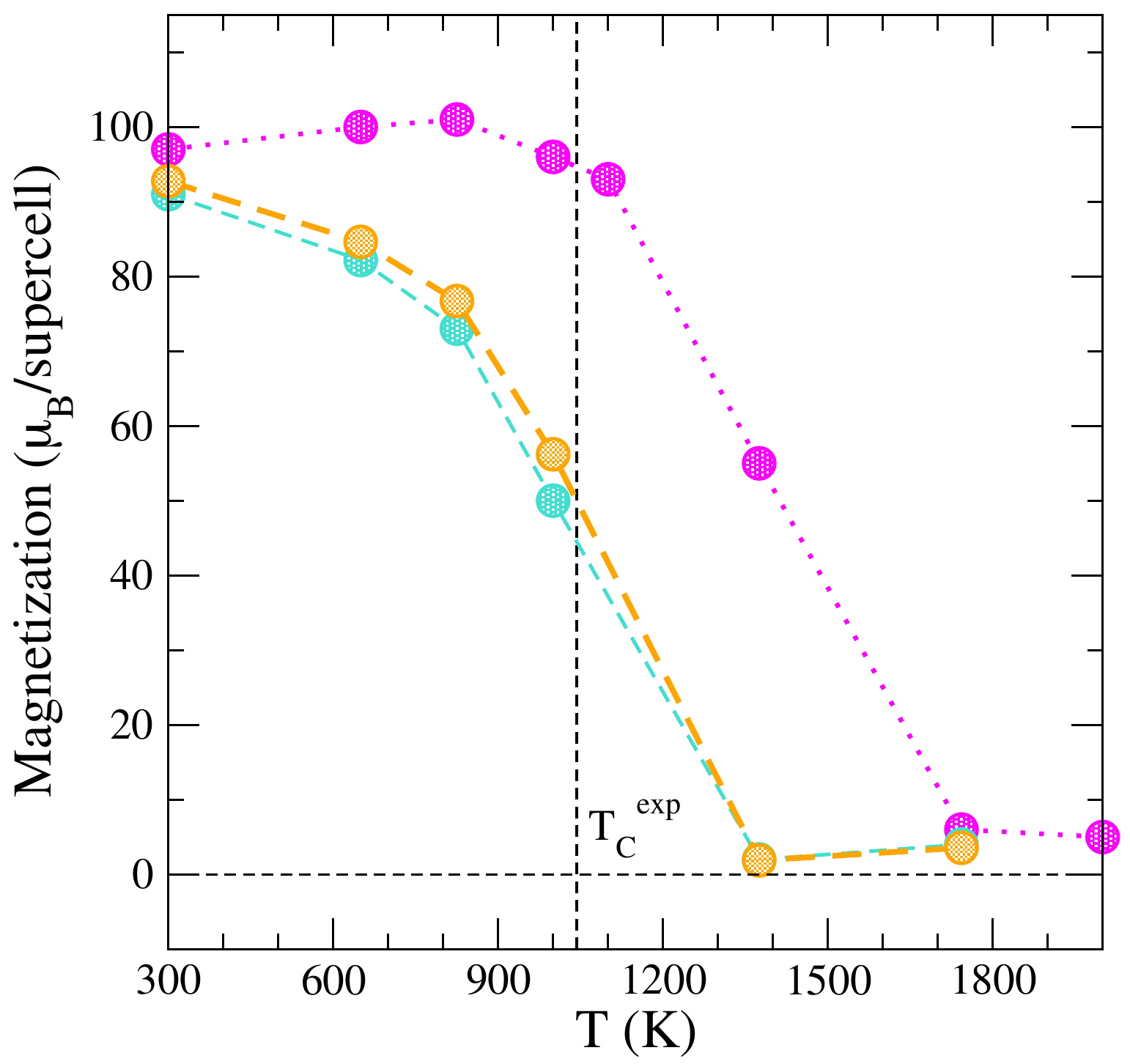}
\caption{Magnetization curve of bcc Fe calculated by the MDMC method for a 54 atom supercell}

\label{fig:Mt_curve}
\end{figure}

Contrary to this behavior, when I applied the MDMC method as described in Sec. \ref{sec:computational} and calculated the time-averaged magnetization, a clear magnetic phase transition was observed, see the purple curve in Fig. \ref{fig:Mt_curve}. 
The transition temperature is close to 1500 K, which is rather large in comparison with the experimental value, though at the same accuracy as in some of the recent calculations \cite{Han}. This result could be further improved by taking into account the fact that 
the average magnetic moment in bcc Fe is expectedly sensitive to the changes in the lattice parameter. The time-averaged pressures in the runs were of order -4 to -8 GPa. That is in line with the known underestimation of the equilibrium volume of Fe in DFT claculations with standard PAW potentials.  
Accordingly, 
I decided to use a second set of lattice parameters to bracket the calculated zero pressure magnetizations 
and run 5 additional points with the smaller lattice constants that resulted in time-averaged pressures between -1 and +3 GPa. The results are presented as a turquoise curve in Fig. \ref{fig:Mt_curve}. Then a simple linear interpolation based on the time-averaged calculated pressures for both sets of points was used to estimate
the magnetization at equilibrium conditions. This bracketing procedure looks like a reasonable time-saving choice for a high-throughput screening. The results are shown as the orange curve in Fig. \ref{fig:Mt_curve}.
To be on a safe side one can accept that the Curie temperature estimate is somewhere between 1000 and 1375~K. This appears to be in good agreement with experimental data. 
Generally speaking, this should be considered as a successful prediction, as the accuracy is at least not worse than in large supercell calculations \cite{T_C_Fe_bcc_calc_HellsvikPRB2019_largecell}. That should be perfectly sufficient for a high-throughput screening. 
As those MDMC runs were done for quite long simulation times (see Sec. \ref{sec:computational}) to make sure the results are well converged, a natural question is how long should be the runs to get the results acceptable for an initial high-throughput screening.

\begin{figure}
\centering

\includegraphics[width=0.48\textwidth]{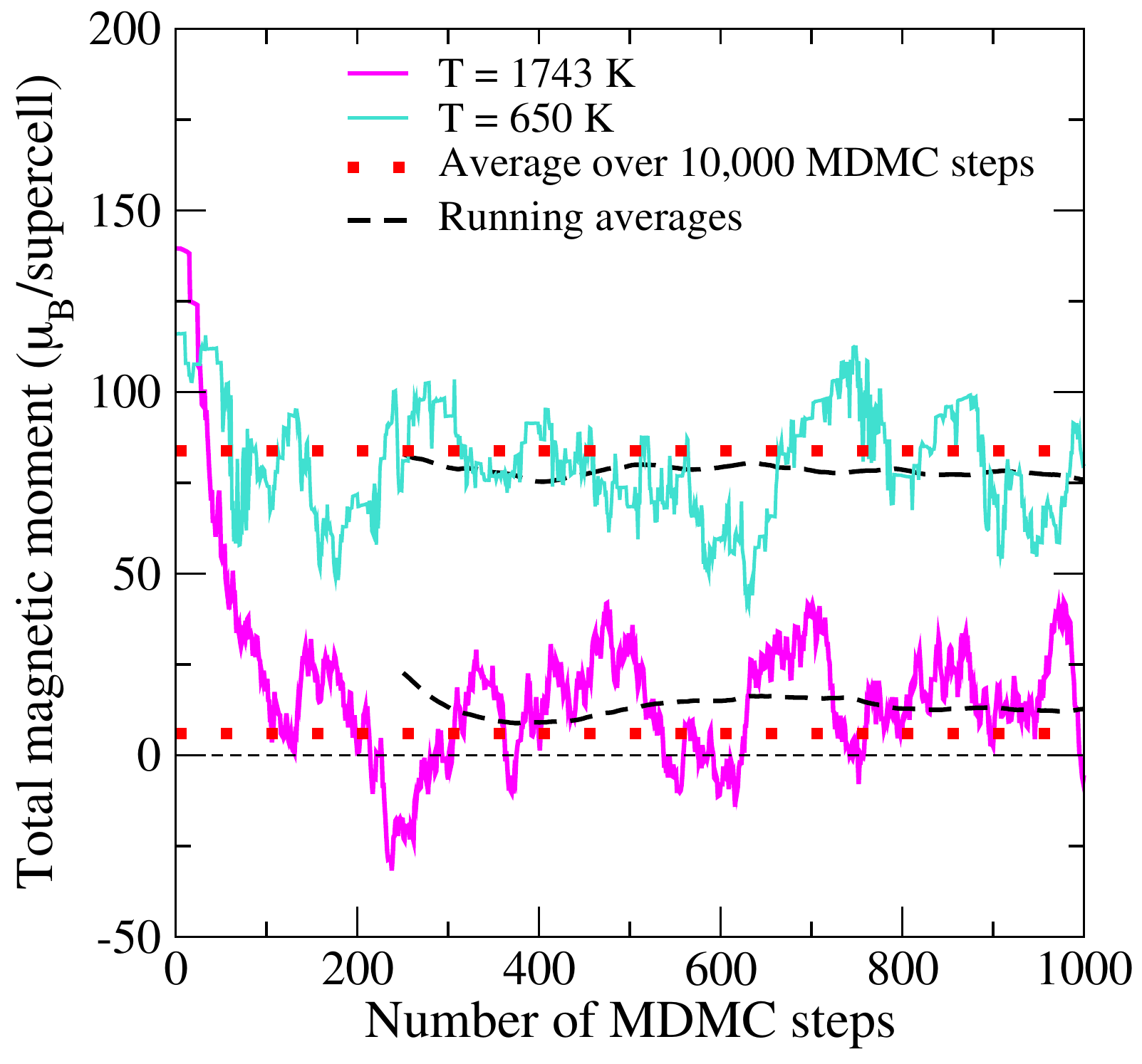}
\caption{The dependence of magnetic moment on the MDMC step obtained by the combined MDMC routine for the bcc Fe at 650 K, below Curie temperature, (shown with purple) and 1743 K, above Curie temperature  (shown with turquoise). Average values are shown with red dotted lines for both temperatures. Running averages are shown with black dashed lines}

\label{fig:MD1000}
\end{figure}

In Fig. \ref{fig:MD1000} one can see the typical MDMC dependence of the magnetic moment of bcc Fe simulated at 650~K and 1743~K at the points on the turquoise and purple curves in Fig. \ref{fig:Mt_curve}, respectively. That is these simulations are done for the temperature below the ferromagnetic-to-paramagnetic transition and in the high-temperature paramagnetic state, also known as $\delta$-phase of Fe, respectively. To notice, these are also the temperatures, for which large amount of different experimental data is available 
\cite{Fe_bcc_exp_highT_phonons_elasticconstants}. 

The magnetic moment in Fig. \ref{fig:MD1000} is plotted versus the number of MDMC steps. 
It is clear from Fig. \ref{fig:MD1000} that the system at  650~K stays in the FM state, 
though magnetic moment flips do occur in the course of the MDMC simulation. However, the system at 1743~K quickly evolves from the FM into the DLM state. Though the on-site magnetic moments show no sign of dying out, the time-averaged total magnetization is close to zero. 
It is important to notice that due to the coupling to atomic vibrations equilibration of the
magnetic configuration does not take longer simulation time than the standard MD equilibration
itself. As may be seen from Fig. \ref{fig:MD1000} for this rather small supercell just around 250
MDMC steps turn out to be sufficient for the system to reach equilibrium.

In order to properly estimate the magnetic moment of the system at a given temperature one needs to average over sufficiently large number of time-steps after the equilibration. I notice, however, that for a quick estimate, 
for example, to discard low-T$_C$ ferromagnets, a relatively short run may be sufficient. Indeed,  using a rather small supercell for just 1000~fs appear to be a good choice in the case of high-temperature bcc Fe.

The averages shown with red dotted lines in Fig. \ref{fig:MD1000} are calculated for the run of 10 ps or 10,000 MDMC steps. I also show running averages to see the tendency to convergence (with black dashed line). As one can see from Fig. \ref{fig:MD1000}, the magnetic moment of the system at 650~K fluctuates around $\sim$ 84~$\mu$B per supercell, while the magnetic moment of the system at 1743~K fluctuates around almost zero. This certainly means that these two points represent two different magnetic states of the system. Therefore, I suggest that short MDMC runs of order 1000 MDMC steps on reasonably small supercells may be helpful if incorporated in high-throughput screening of magnetic materials at finite temperatures.

\section*{\label{sec:conclusions}Conclusions}
I have introduced MDMC, the new algorithm allowing to treat atomic vibrations and magnetic ordering simultaneously in a framework of combined ab initio molecular dynamics and Monte Carlo simulations. At zero K temperature MDMC finds the true magnetic ground state of the system without a priori information from the literature. The traditionally used theoretical approach is the comparison of the equilibrium energies of two competing magnetic states, such as FM and the simplest AFM orderings, in order to find the most energetically stable one. As sometimes the AFM states can be of a rather complex type, such calculations may lead to wrong conclusions.  
However, the offered here approach allows one to quickly and efficiently find the solution in one run.

I believe one cannot claim the structural stability of a system at given conditions without the proper information on its magnetic state. This means that along with the energy determination through a self-consistent DFT cycle one needs to have another self-consistent magnetic cycle to obtain the proper magnetic state. 
The suggested technique provides such a possibility. 
It might be especially helpful for treating compounds with antiferromagnetic, ferrimagnetic and other complex distributions of magnetic moments including the  non-collinear ones. It also allows one to relatively quickly find the equilibrium temperature-dependent magnetic state, which may have dynamic nature, and estimate the magnetic transition temperatures.

I have shown how MDMC can successfully find the ferromagnetic ground state of bcc Fe in a supercell geometry if the initial magnetic distribution is not ferromagnetic, though a standard  DFT calculation would fail to do so. MDMC is also able to predict the very complex magnetic state of $\alpha$-Mn with different values and directions of magnetic moments starting from a completely wrong initial magnetic structure. Further,  MDMC has revealed that the assumption that MnB$_2$W$_2$ is non-magnetic in its ground state, which has been persisting in the literature for 50 years and has been recently confirmed theoretically, is incorrect. For the first time it has been classified as A-type antiferromagnet. This illustrates that the method also works for complex compounds. Finally, I have shown that MDMC works really well in the case of high temperature FM-PM transition in bcc Fe, both showing the proper magnetic ordering depending on temperature and the estimated Curie temperature.
I expect that the MDMC method might be especially suitable for high-throughput screenings of magnetic materials, such as hard magnets or magnetocalorics. The method has been developed into a user-friendly code with multiple options, which is available on a reasonable request.

\begin{acknowledgments}
The author would like to acknowledge the support of  Sweden's Innovation Agency (Vinnova), Equal Opportunities grant from Uppsala University, and the Swedish Research Council (VR) grant \#2019-06063. The computations were performed on resources provided by the Swedish National Infrastructure for Computing (SNIC) at PDC and NSC centers.

\end{acknowledgments}

\bibliographystyle{apsrev4-1}
\bibliography{references.bib}

\end{document}